\documentclass[lettersize,journal]{IEEEtran}
\usepackage{amsmath,amsfonts,amssymb}
\usepackage{algorithmic}
\usepackage[ruled,linesnumbered]{algorithm2e}
\usepackage{array}
\usepackage[caption=false,font=normalsize,labelfont=sf,textfont=sf]{subfig}
\usepackage{textcomp}
\usepackage{stfloats}
\usepackage{url}
\usepackage{verbatim}
\usepackage{graphicx}
\usepackage{cite}
\usepackage{hyperref}
\usepackage{bm} 
\usepackage{upgreek}
\usepackage[normalem]{ulem}
\usepackage{multirow}
\makeatletter

\newcommand{\Rmnum}[1]{\expandafter\@slowromancap\romannumeral #1@}
\makeatother

\begin{document}

\title{Wavelet-Optimized Motion Artifact Correction in 3D MRI Using Pre-trained 2D Score Priors}

\author{Genyuan Zhang, Xuyang Duan, Songtao Zhu, Ao Wang, Fenglin Liu
\thanks{This work was supported by the National Key Research and Development Pro-gram of China (No. 2022YFF0706400), the National Natural Science Foundation of China (N0. 62172067), and the Fundamental Research Funds for the Central Universities (No.2024CDJYXTD-009). F. Liu is the corresponding author.
G. Zhang, X. Duan, S. Zhu, A. Wang, and F. Liu are with the Key Lab of Optoelectronic Technology and Systems, and Engineering Research Center of Industrial Computed Tomography Nondestructive Testing, Ministry of Education, Chongqing University, Chongqing, China, 400044. (zhanggy@stu.cqu.edu.cn, 202408021042T@stu.cqu.edu.cn, 202408131056T@stu.cqu.edu.cn, wao@stu.cqu.edu.cn and liufl@cqu.edu.cn) 

}
}

\markboth{Journal of \LaTeX\ Class Files,~Vol.~14, No.~8, August~2021}%
{Shell \MakeLowercase{\textit{et al.}}: A Sample Article Using IEEEtran.cls for IEEE Journals}


\maketitle

\begin{abstract}
Motion artifacts in magnetic resonance imaging (MRI) remain a major challenge, as they degrade image quality and compromise diagnostic reliability. Score-based generative models (SGMs) have recently shown promise for artifact removal. However, existing 3D SGM-based approaches are limited in two key aspects: (1) their strong dependence on known forward operators makes them ineffective for correcting MRI motion artifacts, and (2) their slow inference speed hinders clinical translation. To overcome these challenges, we propose a wavelet-optimized end-to-end framework for 3D MRI motion correct using pre-trained 2D score priors (3D-WMoCo). Specifically, two orthogonal 2D score priors are leveraged to guide the 3D distribution prior, while a mean-reverting stochastic differential equation (SDE) is employed to model the restoration process of motion-corrupted 3D volumes to motion-free 3D distribution. Furthermore, wavelet diffusion is introduced to accelerate inference, and wavelet convolution is applied to enhance feature extraction. We validate the effectiveness of our approach through both simulated motion artifact experiments and real-world clinical motion artifact correction tests. The proposed method achieves robust performance improvements over existing techniques. Implementation details and source code are available at: https://github.com/ZG-yuan/3D-WMoCo.
\end{abstract}


\section{Introduction}
\IEEEPARstart{M}{AGNETIC} resonance imaging (MRI) is a non-invasive modality that has become an essential diagnostic tool for a wide range of clinical conditions, such as brain tumors \cite{despotovic2015mri}, Alzheimer's disease \cite{jack2015magnetic} and cerebrovascular disease \cite{kidwell2006imaging}. Nevertheless, the scanning time of MRI is relatively long, so motion artifacts caused by physiological factors such as breathing of patients are often unavoidable \cite{zaitsev2015motion}. This motion artifact is particularly pronounced in certain patient populations, such as infants \cite{yoshida2013diffusion}, patients with movement disorders \cite{mascalchi2012movement}, or patients with psychological disorders \cite{agarwal2010update}, who often struggle to maintain immobility during the MRI procedure. Consequently, developing robust methods for correcting MRI motion artifacts is of great importance for improving diagnostic reliability and patient care.

Numerous approaches have been explored for motion artifact mitigation.
Compressed sensing (CS)-based approaches have also gained traction for addressing motion-related artifacts \cite{vasanawala2010improved, yang2013sparse}. However, their reliance on hand-crafted priors and high computational complexity hinders their generalizability across diverse clinical settings. In contrast to traditional approaches, learning-based motion correction (MoCo) methodologies have demonstrated superior performance in addressing MRI motion artifacts, encompassing both image-domain based methods \cite{levac2022fse, tamada2020motion, pawar2020clinical,liu2020motion,lee2021mc2} and k-space based methods \cite{singh2024data,levac2024accelerated,eichhorn2023physics}. Recently, score-based generative models (SGMs) have emerged as a promising paradigm, as they can leverage the distribution of artifact-free data for motion correction \cite{spieker2023deep, oh2023annealed, safari2024mri}.

\begin{figure}[!t]
\centerline{\includegraphics[width=\linewidth]{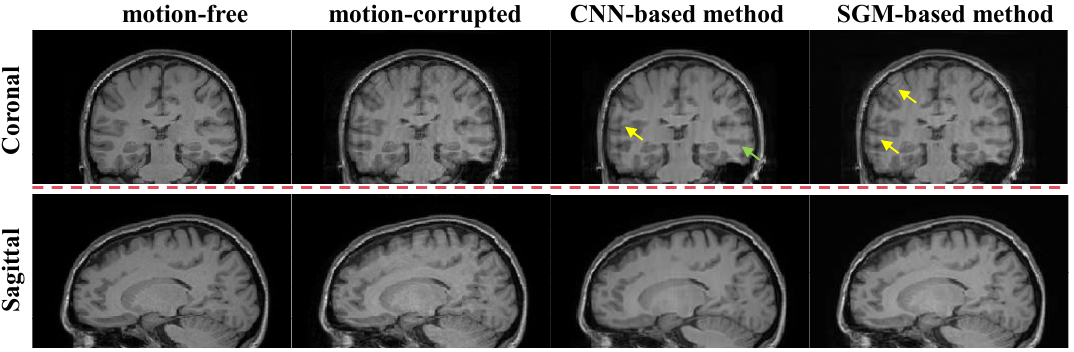}}
\caption{Results of CNN-based and SGM-based motion correction methods trained on the sagittal plane. Yellow arrows highlight residual motion artifacts, while green arrows indicate inter-slice inconsistencies.}
\label{fig4}
\end{figure}

Most existing methods remain limited to two-dimensional (2D) motion correction. However, motion artifacts have different artifact characteristics and data distributions in different anatomical sections \cite{duffy2021retrospective}. 2D-based methods have difficulty in fully capturing the features of all anatomical sections, resulting in obvious residual motion artifacts in non-training planes, a limitation clearly illustrated by the yellow arrow in Fig. \ref{fig4}. Furthermore, stacking independently reconstructed 2D slices to form a three-dimensional (3D) volume lead to inter-slice discontinuities, as shown by the green arrow in the fig. \ref{fig4} \cite{reyneke2018review}. These limitations impair the accuracy and coherence of 3D volumetric reconstructions, posing significant challenges for clinical applications that rely on consistent anatomical representations \cite{mediouni2018retracted, mediouni2020translational}.

A straightforward solution is to directly train 3D diffusion models, but this strategy faces three major challenges. First, 3D diffusion models impose substantial computational demands during inference, requiring high-performance GPU resources \cite{li2024two}. Second, training a 3D model requires thousands of volumetric datasets, whereas a 2D model can be trained with only dozens \cite{chung2023solving}. Finally, directly training 3D diffusion would unnecessarily complicate the prior knowledge \cite{jung2009k, sun2019exploiting}. To alleviate these issues, prior studies have proposed some strategies: For example, Chung et al. \cite{chung2023solving} incorporated total variation (TV) regularization along the Z-axis to enhance 2D score models, while Lee et al. \cite{lee2023improving} trained two orthogonal 2D score models to capture inter-slice information. However, these approaches share a fundamental limitation: they necessitate explicit knowledge of the 3D image degradation process, which can be mathematically represented as: ${\bf{y}} = {\bf{Ax}} + {\bf{n}}$,
where $({\bf{x}}, {\bf{y}})$ are high-quality and low-quality images, respectively, ${\bf{A}}$ represents an explicit degradation mechanism, and ${\bf{n}}$ is noise.  In real-world medical imaging scenarios, particularly for tasks like MRI motion artifact removal, accurately defining the degradation process ${\bf{A}}$ proves elusive due to the inherent complexity and variability of motion patterns. Each unique motion pattern often requires tailored modeling strategies, significantly escalating the complexity of the overall model architecture. Moreover, computational efficiency poses another critical challenge. Moreover, existing methods often require more than ten hours to reconstruct a single 3D volume of size $(256\times256\times256)$, which is impractical for clinical workflows. 

To address the challenges in existing methods, we should directly establish the restoration process of motion-corrupted 3D volumes to motion-free 3D distribution. Mean-Reverting stochastic differential  equations \cite{luo2023image} provides a potential solution, as they restore low-quality images by transforming high-quality images into a mean state with Gaussian noise, without relying on task-specific prior knowledge. Furthermore, the 3D distribution can be approximated from pre-trained 2D score models, thereby avoiding the computational and data burdens of direct 3D training. Although the reasoning speed in \cite{luo2023image} has been greatly improved due to the reduction of diffusion step size, we can still use methods such as wavelet decomposition \cite{zhang2019wavelet} to reduce the dimension of the input image to further enhance the reasoning speed. 

Building on these insights, we propose a wavelet-optimized end-to-end framework for 3D MRI motion-correct using pre-trained 2D score priors (3D-WMoCo). Specifically, we generalize mean reversion SDE to 3D and directly model the motion corruption process from motion-free MRI volumes to motion-corrupted volumes. This eliminates the need for the model to understand the exact degradation process or motion pattern, improving the practicality of the model. The score required for 3D mean-reverting SDE is derived from two orthogonal 2D scores, avoiding the problem of training 3D scores while maintaining critical volume coherence. In addition, we perform diffusion in the wavelet domain for faster processing while maintaining good generation quality. Finally, we replace ordinary convolutions in the model with wavelet convolutions to expand the receptive field and better capture low-frequency signals \cite{finder2024wavelet}. This paper presents four significant contributions, which can be summarized as follows:
\begin{itemize}
\item
We generalize mean regressive SDE to 3D and directly model the motion corruption process. This allows the model to learn a general degradation manifold and generalize to a variety of motion artifacts.
\end{itemize}
\begin{itemize}
\item
We trained two orthogonal 2D scores for repairing 3D MRI motion-damaged volumes, avoiding the high computational and data requirements of full 3D training.
\end{itemize}
\begin{itemize}
\item
We further accelerate the inference speed through wavelet diffusion and use wavelet convolution to expand the receptive field to extract features more finely.
\end{itemize}
\begin{itemize}
\item
We achieve highly competitive performance on both simulated and real 3D motion corruption data.
\end{itemize}

\begin{figure*}
\centerline{\includegraphics[width=\textwidth]{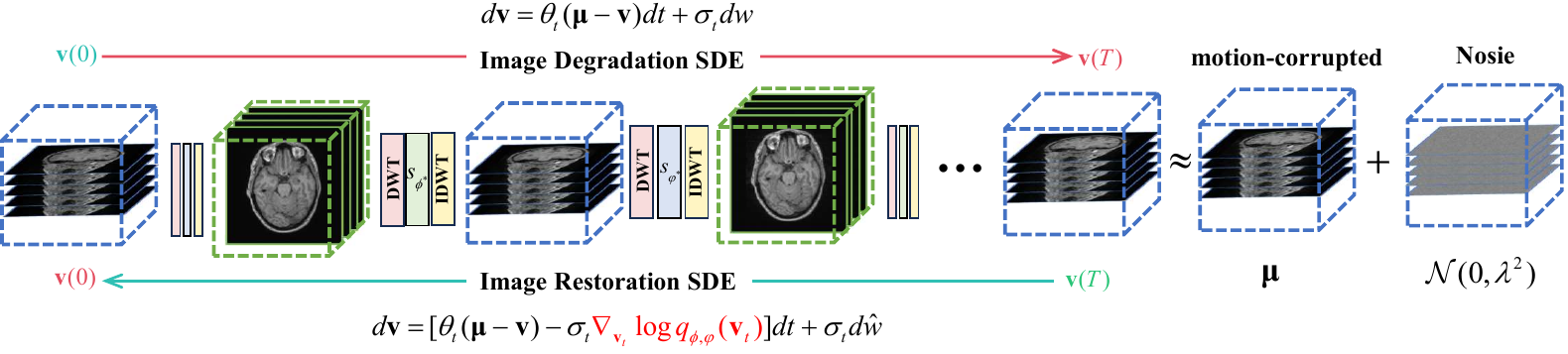}}
\caption{Overview of the Proposed 3D-WMoCo. The diffusion endpoint of the proposed model is a motion-corrupted volume ${\bf{\mu }}$ with added Gaussian noise ${\cal N}(0,{\lambda ^2})$. We perform the restoration process from a motion-corrupted MRI volume ${\bf{v}}(T)$ to a motion-free MRI volume ${\bf{v}}(0)$ by alternately using the scores obtained from two orthogonal slices ${s_{{\phi ^ * }}}$ and ${s_{{\varphi ^ * }}}$. The model is accelerated and optimized by incorporating wavelet transforms.}
\label{fig1}
\end{figure*}

\section{Related Works}
\subsection{Deep Learning for 3D MRI Motion Artifact Correction}
Motion artifacts in medical imaging arise as spatially correlated distortions across 3D volumes, necessitating methodologies that explicitly model volumetric contextual information for effective mitigation. With the development of deep learning, some methods have emerged to solve 3D motion artifacts through deep learning. Convolutional neural networks (CNNs), with their hierarchical feature extraction capabilities, have been pivotal in this domain. For instance, Lv et al. \cite{lv2018respiratory} pioneered a 3D CNN-based registration method for free-breathing abdominal MRI. Pirkl et al. \cite{pirkl2022learning} employed a 3D multi-scale CNN to learn the residual relationship between motion artifacts and artifact-free images. Miller et al. \cite{miller2023motion} integrated motion compensation with a multi-scale CNN to estimate motion fields. Generative adversarial networks (GANs) have also emerged as effective tools for modeling complex motion artifact distributions. For example, Johnson et al. \cite{johnson2019conditional} utilized conditional GANs for 3D rigid motion correction, while Ghodrati \cite{ghodrati2021temporally} combined spatiotemporal correlation with adversarial training to achieve 3D respiratory motion compensation. However, these approaches often result in high GPU consumption. Moreover, compared with SGM, CNN/GAN methods exhibit limited capability in modeling complex motion and generalizing to unknown motion models.
\subsection{Mean-Reverting Stochastic Differential Equations}
Given a initial state ${{\bf{x}}_0}$ sampled from an unknown distribution, ${{\bf{x}}_0} \sim {p_0}({\bf{x}})$, the forward mean-reverting SDE \cite{luo2023image} is defined according to:
\begin{equation} 
d{\bf{x}} = {\theta _t}({\bf{\mu }} - {\bf{x}})dt + {\sigma _t}dw,\label{eq22}\end{equation}
where ${\bf{\mu }}$ is the state mean, ${\theta _t}$ and ${\sigma _t}$ are time-varying positive parameters that characterize the mean reversion speed and stochastic volatility, respectively, and $w$ is a standard Wiener process. To recover the high-quality image from the terminal state ${\bf{x}}(T)$. Reversing equation \ref{eq22}, we get the image recovery SDE (IR-SDE), which is given by:
\begin{equation} 
d{\bf{x}} = [{\theta _t}({\bf{\mu }} - {\bf{x}}) - \sigma _t^2{\nabla _{\bf{x}}}\log {q_t}({{\bf{x}}_t})]dt + {\sigma _t}d\hat w,\label{eq23}\end{equation}
where ${\nabla _{\bf{x}}}\log {q_t}({{\bf{x}}_t})$ is the score function and this score can be estimated by training a neural network ${{{\tilde \epsilon}_\phi }}$. 
Then, we optimize the network to make the reverse IR-SDE optimal:
\begin{equation} 
J_\gamma(\phi):=\sum_{i=1}^T \gamma_i \mathbb{E}[\|\underbrace{x_i-\left(\mathrm{d} x_i\right)_{{{\tilde \epsilon}_\phi }}}_{\text {reversed } x_{i-1}}-x_{i-1}^*\|]
,\label{eq24}\end{equation}
where ${\left( {{\rm{d}}{x_i}} \right)_{{{\tilde \epsilon }_\phi }}}$ denoted the reverse-time SDE in Eq. \ref{eq23} and its score is predicted by the noise network ${{{\tilde \epsilon }_\phi }}$. At last, for a given initial state ${{\bf{x}}_0}$, the optimal inversion solution ${\bf{x}}_{t - 1}^ *$ can be given by the following equation: 

\begin{equation} 
\begin{array}{l}
{\bf{x}}_{t - 1}^ *  = \frac{{1 - {e^{ - 2{{\overline \theta  }_{t - 1}}}}}}{{1 - {e^{ - 2{{\overline \theta  }_t}}}}}{e^{ - {{\theta '}_t}}}\left( {{{\bf{x}}_t} - \bm{\upmu} } \right)\\
 + \frac{{1 - {e^{ - 2{{\theta '}_t}}}}}{{1 - {e^{ - 2{{\overline \theta  }_t}}}}}{e^{ - {{\overline \theta  }_{t - 1}}}}\left( {{{\bf{x}}_0} - \bm{\upmu} } \right) + \bm{\upmu}
\end{array},\label{eq24}\end{equation}
where ${{\theta '}_t} = \int_{i - 1}^i {{\theta _t}} dt, {{\bar \theta }_t} = \int_0^t {{\theta _j}} dj$. 

\begin{algorithm}
    \SetAlgoLined 
    \caption{3D-WMoCo: Training and Sampling Process} 
    \KwIn{Motion-corrupted/motion-free MRI images set ${\bf{I}} = \{ {({{\bf{v}}_0},{{\bf{v}}_T})_i}\} _{i = 1}^N, {\bf{v}} \in \mathbb{R}{^{d1,d2,d3}}$, total time steps $T$} 
    \KwOut{De-motion artifact image: ${{\bf{v}}_{0}}$} 
    \textbf{\uppercase\expandafter{\romannumeral1}. Training Process} \;
    Sample $DWT({{\bf{v}}_{0,[:,:,i]}},{{\bf{v}}_{T,[:,:,i]}})_{i = 1}^{d3}$\;
    According to Eq. \ref{eq24}, the score ${s_\phi }$ of a face is obtained \;
    Sample  $DWT({{\bf{v}}_{0,[:,i,:]}},{{\bf{v}}_{T,[:,i,:]}})_{i = 1}^{d2}$\;
    According to Eq. \ref{eq24}, the score ${s_\varphi }$ of a face is obtained \;
    \textbf{\uppercase\expandafter{\romannumeral2}. Sampling Process} \;
    Data initialization: $\bf{W} = zeros\_like({{\bf{v}}_T})$\;
    \For{$t = T$ \textbf{down to} $0$}{
        \eIf{mod(t,2)=0}{
        \For{$i = 1$ \textbf{up to} $d3$}{
            ${{\bf{v'}}} = squeeze({{\bf{v}}_T}_{,[:,:,i]})$,
            $\bm{\upmu}  = squeeze({{\bf{v}}_T}_{,[:,:,i]})$,
            ${{\bf{v'}}_t}\mathop  \leftarrow \limits^{add\_t\_noise} {\bf{v'}}$ \;
            \eIf{t=T}{
                ${{\bf{W}}_{[:,:,i]}}\mathop  \leftarrow \limits^{add\_T\_noise} {\bf{v'}}$ \;
            }
            {
                ${\bf{v'}}_t = {{\bf{W}}_{[:,:,i]}}$, $v_t^{wav} = DWT({{\bf{v}}_t})$ \;
                According to Eq. \ref{eq4}, ${\bf{v'}}_{t-1}^{wav}$ is obtained \;
                ${{\bf{W}}_{[:,:,i]}} = IDWT({{\bf{v'}}_{t - 1}})$ \;
            }
        }
        }
        {
        \For{$i = 1$ \textbf{up to} $d2$}{
            ${\bf{v'}} = squeeze({{\bf{v}}_T}_{,[:,i,:]})$,
            $\bm{\upmu}  = squeeze({{\bf{V}}_T}_{,[:,i,:]})$,
            ${{\bf{v'}}_t}\mathop  \leftarrow \limits^{add\_t\_noise} {\bf{v'}}$ \;
            \eIf{t=T}{
                ${{\bf{W}}_{[:,i,:]}}\mathop  \leftarrow \limits^{add\_T\_noise} {\bf{v'}}$ \;
            }
            {
            ${\bf{v'}}_t = {{\bf{W}}_{[:,i,:]}}$, ${\bf{v'}}_t^{wav} = DWT({{\bf{v}}_t})$ \;
            According to Eq.\ref{eq4}, ${\bf{v'}}_{t-1}^{wav}$ is obtained \;
            ${{\bf{W}}_{[:,i,:]}} = IDWT({{\bf{v'}}_{t - 1}})$ \;
            }
        }
        }
    }
    \textbf{return} $\bf{v}_{0}$ \;
\end{algorithm}

\begin{figure}[!t]
\centerline{\includegraphics[width=\linewidth]{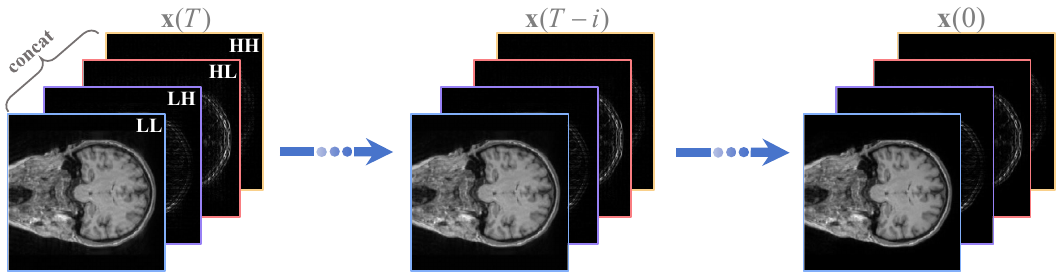}}
\caption{The model is implemented in the wavelet domain. The image is decomposed into four wavelet components, which are then concatenated along their dimensions.}
\label{fig3}
\end{figure}

\section{Method}
\subsection{3D Modeling}
To effectively leverage IR-SDE to solve the problem of 3D MRI motion artifacts, we conduct the training of a two-dimensional (2D) diffusion model along two orthogonal directions. Subsequently, we employ a straightforward yet highly efficient collaborative strategy to eliminate the 3D motion artifacts using 2D a priori information.

We assume that the 3D body distribution is $p(X,Y,Z)$. The conditional distributions of the three orthogonal planes are: $p(X,Y|Z = z)$, $p(X,Z|Y = y)$, $p(Z,Y|X = x)$. The edge distributions of the three axes X, Y, and Z are: $P(X)$, $P(Y)$, $P(Z)$. If we assume that the distribution of each face is conditionally independent, e.g., the distribution of the XY face depends only on the current Z-value and is conditionally independent between slices of different Z-values, the 3D distribution will be expressed as follows:
\begin{equation} \begin{aligned}
p(X,Y,Z) = {p(X,Y|Z )p(Z)}= {p(X,Z|Y)p(Y)} \\
         = {p(Z,Y|X )p(X)} 
\end{aligned}.\label{eq1}\end{equation}
Here, we assume that for a given position along one axis (e.g., the X axis), the pixel values along the other two axes (Y and Z) are approximately independent of each other. We acknowledge that this assumption has certain flaws, but deep learning models can compensate for these deviations through data learning, making the model robust in practice. According to this assumption we have:
\begin{equation} 
P(X,Y|Z) = P(Y|X)P(Z|X)
.\label{eq1122}\end{equation}
Substituting the above formula into Eq.\ref{eq1}, and obtain the following according to the conditional probability formula: 
\begin{equation*} 
\begin{aligned}
p(X,Y,Z) &= P(Y|X)P(Z|X)P(X) \\
 &= \frac{{P(X,Y)}}{{P(X)}} \cdot \frac{{P(X,Z)}}{{P(X)}}\cdot P(X) \\
 &= \frac{{P(X,Y)\cdot P(X,Z)}}{{P(X)}}
\end{aligned}.\label{eq1111}\end{equation*}
We then parametrize the above formula using the following geometric weighted average, where a and b are used to balance the two distributions:
\begin{equation} 
p(X,Y,Z)
 = ({({p(X,Y)} )^\alpha }
 \times {({p(X,Z)})^\beta })/Z, \label{eq2}\end{equation}
where $\alpha$ and $\beta$ are used to balance the two distributions, $Z$ is an appropriate normalizing partition function. 
We use ${q_\phi }({\bf{v}})$ and ${q_\varphi }({\bf{v}})$ to denote the distributions parameterized by slices $(X,Y)$ and $(X,Z)$, respectively, where ${\bf{v}} \in \mathbb{R}{^{d1,d2,d3}}$.
The 3D distribution constructed from these two distributions is ${p_{\varphi ,\phi }}({\bf{v}})$. When we sample unconditionally from a 3D prior distribution, we can use:
\begin{equation} 
\begin{array}{l}
{\nabla _{{{\bf{v}}_t}}}\log {q_{\phi ,\varphi }}({{\bf{v}}_t}) = \alpha {\nabla _{{{\bf{v}}_t}}}\log {q_\phi }({{\bf{v}}_t}) + \beta {\nabla _{{{\bf{v}}_t}}}\log {q_\varphi }({{\bf{v}}_t})\\
 = \alpha \sum\nolimits_{i = 1}^{d3} {{\nabla _{{{\bf{v}}_{t,[:,:,i]}}}}\log {q_\phi }({{\bf{v}}_{t,[:,:,i]}})}  + \\
\beta \sum\nolimits_{i = 1}^{d2} {{\nabla _{{{\bf{v}}_{t,[:,i,:]}}}}\log {q_\varphi }({{\bf{v}}_{t,[:,i,:]}})} \\
\end{array},\label{eq4}\end{equation} 
where ${{{\bf{v}}_{t,[:,:,i]}}}$ and ${{{\bf{v}}_{t,[:,j,:]}}}$ denote the i and j-th xy- and xz- slice of ${{\bf{v}}_t}$, respectively. We noticed that calculating the scores of two faces simultaneously would lead to huge computational complexity and time complexity. We use the score functions of the two faces alternately with a certain probability \cite{lee2023improving}, which is expressed as follows:
\begin{equation} 
{\nabla _{{{\bf{v}}_t}}}\log {q_{\phi ,\varphi }}({{\bf{v}}_t}) = \left\{ \begin{array}{l}
\sum {{\nabla _{{{\bf{v}}_t}}}\log {q_\phi }({{\bf{v}}_{t,[:,:,i]}}),
P = \alpha /(\alpha  + \beta )}\\
\sum {{\nabla _{{{\bf{v}}_t}}}\log {q_\varphi }({{\bf{v}}_{t,[:,i,:]}}),P = \beta /(\alpha  + \beta )}
\end{array} \right.,\label{eq5}\end{equation}
where $\alpha  + \beta  = 1$. Therefore, we conclude that the inverse process of 3D IR-SDE is:
\begin{equation} 
d{\bf{v}} = [{\theta _t}(\bm{\upmu}  - {\bf{v}}) - {\sigma _t^2}{\nabla _{{{\bf{v}}_t}}}\log {q_{\phi ,\varphi }}({{\bf{v}}_t})]dt + {\sigma _t}d\hat w.\label{eq6}\end{equation}

To improve sampling efficiency, Zhang et al. \cite{zhang2024entropy} proposed a posterior sampling method, which we extend to 3D. Specifically, When high-quality 3D volume ${{\bf{v}}_{\bf{0}}}$ is given, this posterior distribution is given by:
\begin{equation} 
p({{\bf{v}}_{t - 1}}|{{\bf{v}}_t},{{\bf{v}}_0}) = {\cal N}({{\bf{v}}_{t - 1}}|{{\tilde {\bm{\upmu}} }_t}({{\bf{v}}_t},{{\bf{v}}_0}),{{\tilde \beta }_t}{\bf{I}}),\label{eq7}\end{equation}
which is a Gaussian with mean and variance given by:
\begin{equation} 
\left\{ \begin{array}{l}
\begin{array}{*{20}{l}}
{{{\bf{\tilde {\bm{\upmu}} }}}_{\bf{t}}}({{\bf{v}}_{\bf{t}}},{{\bf{v}}_{\bf{0}}}) = \\
\sum{\frac{{1 - {e^{ - 2{{\bar \theta }_{t - 1}}}}}}{{1 - {e^{ - 2{{\bar \theta }_t}}}}}{e^{ - {{\theta '}_t}}}({{\bf{v}}_{t,[:,:,i]}} - {\bm{\upmu}_{[:,:,i]}}) + }\\
\begin{array}{l}
\frac{{1 - {e^{ - 2{{\theta '}_t}}}}}{{1 - {e^{ - 2{{\bar \theta }_t}}}}}{e^{ - {{\bar \theta }_{t - 1}}}}({{\bf{v}}_{0,[:,:,i]}} - {{\bm{\upmu}}_{[:,:,i]}}) + {\bm{\upmu}_{[:,:,i]}},\\
P = \alpha /(\alpha  + \beta )
\end{array}
\end{array}\\
\begin{array}{*{20}{l}}
{{{{\bf{\tilde {\bm{\upmu}} }}}_{\bf{t}}}({{\bf{v}}_{\bf{t}}},{{\bf{v}}_{\bf{0}}}) = }\\
\sum{\frac{{1 - {e^{ - 2{{\bar \theta }_{t - 1}}}}}}{{1 - {e^{ - 2{{\bar \theta }_t}}}}}{e^{ - {{\theta '}_t}}}({{\bf{v}}_{t,[:,i:]}} - {\bm{\upmu}_{[:,i,:]}}) + }\\
\begin{array}{l}
\frac{{1 - {e^{ - 2{{\theta '}_t}}}}}{{1 - {e^{ - 2{{\bar \theta }_t}}}}}{e^{ - {{\bar \theta }_{t - 1}}}}({{\bf{v}}_{0,[:,i,:]}} - {\bm{\upmu}_{[:,i,:]}}) + {\bm{\upmu}_{[:,i,:]}},\\
P = \beta /(\alpha  + \beta )
\end{array}
\end{array}
\end{array} \right..\label{eq8}\end{equation}
and
\begin{equation} 
{\tilde \beta }_t = \frac{{(1 - {e^{ - 2{{\bar \theta }_{t - 1}}}})(1 - {e^{ - 2{{\theta '}_t}}})}}{{1 - {e^{ - 2{{\bar \theta }_t}}}}}
.\label{eq9}\end{equation}
Combining the reparameterization trick and the 3D noise prediction network ${{\tilde s}_{\phi ,\varphi }}({{\bf{v}}_t},\bm{\upmu},t)$, a method for estimating ${{\bf{v}}_0}$ at time t is provided:
\begin{equation} 
{{\bf{\hat v}}_0} = {e^{{{\bar \theta }_t}}}({{\bf{v}}_t} - \bm{\upmu} - \sqrt {{v_t}} {{\tilde s}_{\phi^* ,\varphi^* }}({{\bf{v}}_t},\bm{\upmu},t) + \bm{\upmu}
.\label{eq10}\end{equation}
Then we iteratively sample reverse states based on this posterior distribution starting from 3D noisy LQ images for efficient restoration. We present the pseudo code of the proposed method in Algorithm 1.

\subsection{Wavelet Acceleration Strategy}
Wavelet transform is a classic image compression technique used to separate different frequency components from the original image. Among them, Haar wavelet transform is widely used in practical applications due to its simplicity \cite{zhang2019wavelet}.  It involves operations: discrete wavelet transform (DWT) and the inverse discrete wavelet transform (IDWT). 

We decompose the input image $\bf{x} \in {\mathbb{R}^{H \times L}} $ into four wavelet subbands: LL, HL, LH, and HH via the DWT. 
As shown in Fig. \ref {fig1}, the wavelet domain components are fed into the network, the score function is obtained in the wavelet domain, and then converted to the image domain for continuous iteration. The above process is equivalent to performing a diffusion process in the wavelet domain, as shown in Fig \ref{fig3}. Benefiting from the reduction of input image resolution, the testing time of the proposed model is reduced. 

\begin{figure}[!t]
\centerline{\includegraphics[width=0.7\linewidth]{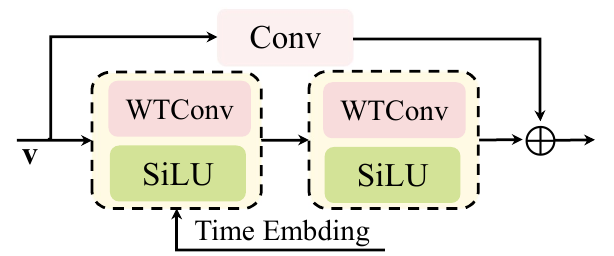}}
\caption{The basic structure of the wavelet residual block is illustrated. Conv denotes the basic convolution block, WTConv is the wavelet convolution block, and SiLU is the activation function.}
\label{fig2}
\end{figure}

\begin{figure*}
\centerline{\includegraphics[width=\textwidth]{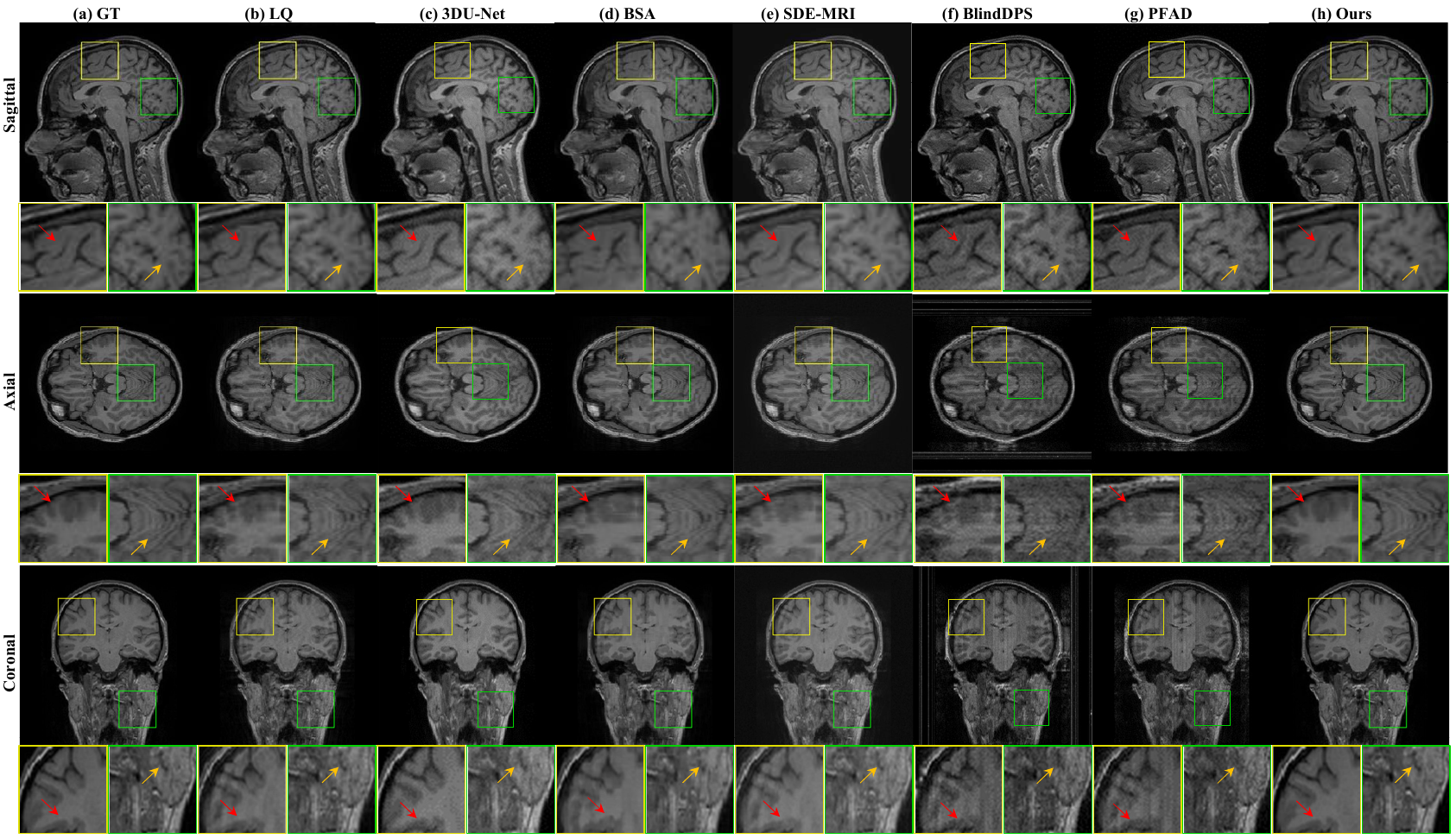}}
\caption{Qualitative results of mild motion artifact correction for brain data in sagittal, coronal, and horizontal plane views using various methods. Yellow and green ROIs are presented as local magnifications. The first column denotes the ground truth (GT), the second column denotes the low-quality image (LQ) corrupted by motion artifacts, the third to sixth columns denote results of comparison algorithms, and the last column denotes our proposed method.}
\label{fig5}
\end{figure*}

\subsection{Wavelet Residual Block}
We use wavelet convolution \cite{finder2024wavelet} to replace the convolutional layer in the traditional residual block, as shown in Fig. \ref{fig2}. For the input tensor ${\bf{x}}$, wavelet convolution can be expressed as follows:
\begin {equation} {\bf {y}} = \text {IDWT}(\text {Conv}({\bf {w}}, \text {DWT}({\bf {x}}))),\label {eq10}\end {equation}
where ${\bf{w}}$ is the weight tensor of a $k  \times k$ depth-wise kernel with four times as many input channels as ${\bf{x}}$. This operation allows the smaller kernel to act on a larger region of the original input, which is equivalent to increasing its receptive field \cite {finder2024wavelet}. Furthermore, we can expand the receptive field through cascading. For a wavelet convolution with an $\ell$-layer, the number of input channels is $c$, and the convolution kernel size is fixed to $k$, the number of parameters grows linearly ($\ell \cdot 4 \cdot c \cdot k$), while the receptive field grows exponentially (${2^\ell } \cdot k$). In addition, the low-frequency information in MRI data can be better captured by incorporating wavelet convolution. Considering the amount of computation, we finally opt for two layers of wavelet convolutions to replace the convolution in the original residual block.

\begin{figure*}
\centerline{\includegraphics[width=\textwidth]{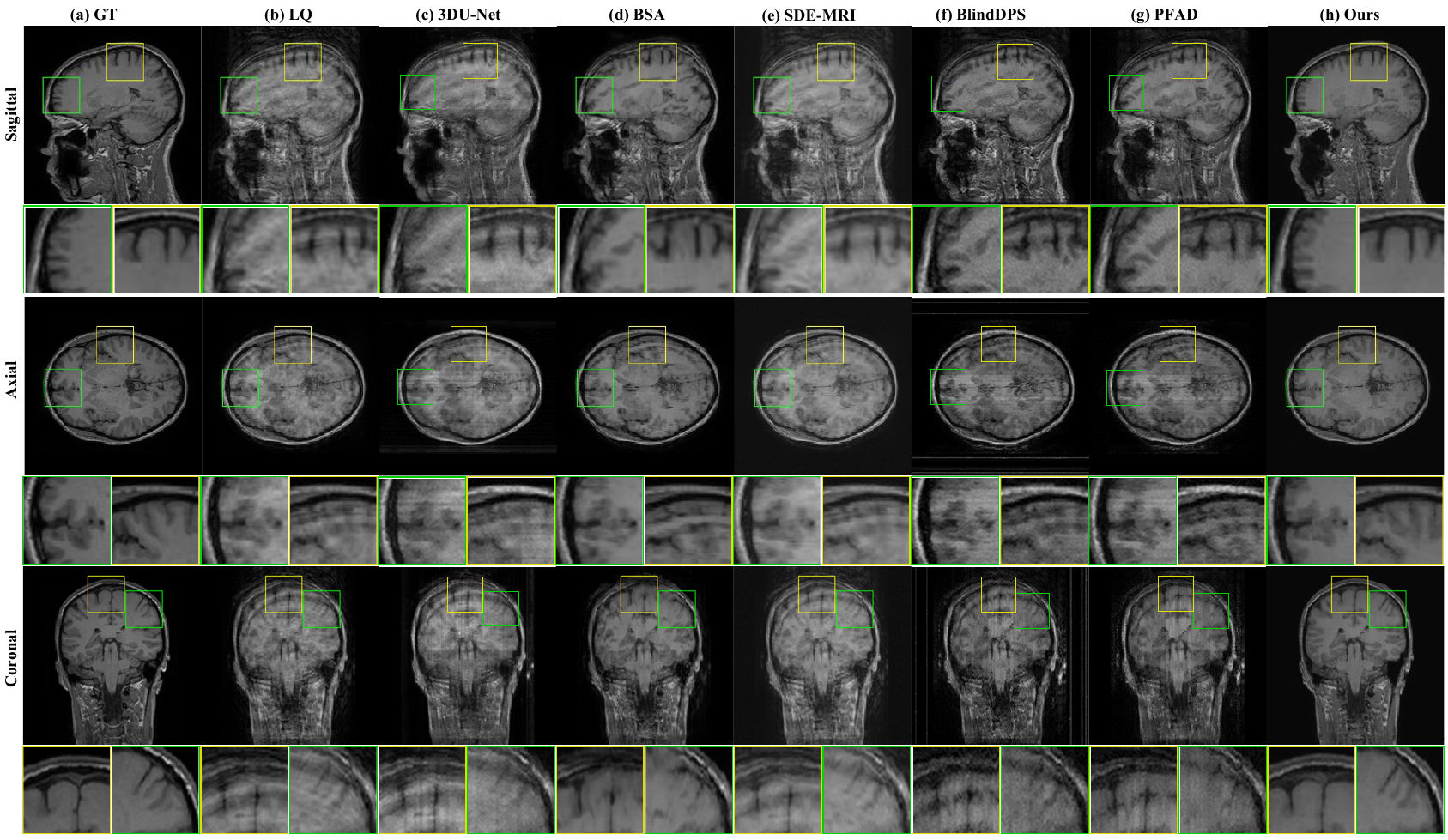}}
\caption{Qualitative results of severe motion artifact correction for brain data in sagittal, coronal, and horizontal plane views using various methods. Yellow and green ROIs are presented as local magnifications. The first column denotes the ground truth (GT), the second column denotes the low-quality image (LQ) corrupted by motion artifacts, the third to sixth columns denote results of comparison algorithms, and the last column denotes our proposed method.}
\label{fig6}
\end{figure*}

\section{Experiments}
\subsection{Datasets}
Three-dimensional T1-weighted MRI volumes from the IXI dataset (\href{http://brain-development.org/ixi-dataset/}{https://brain-development.org/ixi-dataset/}) were used for training and testing our model. 35 subjects were randomly selected from the dataset, of which 31 subjects ($240 \times240 \times240 $) were used as the training set and 4 subjects were used as the test set. According to our evaluation the dataset were considered to be motion-free data. We simulated motion artifacts using the method proposed by \cite{duffy2021retrospective} (\href{https://github.com/bduffy0/motion-correction}{https://github.com/bduffy0/motion-correction}). This method performs a 3D fast Fourier transform (3DFFT) on the 3D MRI images and randomly perturbs the Fourier domain to simulate various head movements. The parameters ${M_{\max }}$ and ${M_{\min }}$ are used to control the percentage range of Fourier lines affected by motion to simulate different degrees of motion artifacts. We simulated two different degrees of motion artifacts, ${M_{\min }} = 30\% ,{M_{\max }} = 45\% $ and ${M_{\min }} = 45\% ,{M_{\max }} = 50\% $, representing mild motion artifacts and severe motion artifacts respectively.

To test the effect of our method on actual motion artifacts, we used the MR-ART dataset \cite{narai2022movement}, consisting of data from 20 healthy adults, which includes both motion-free and different levels of motion-affected data acquired from the same subjects. However these data are not fully paired. The motion-free data are regarded as the ground truth, and the motion-affected data are used as the input of the model.

\subsection{Implementation Details}
We implemented 3D-WMoCo in PyTorch and trained it on one NVIDIA RTX 4090 GPU (24GB) with a batch size of 4. A cosine noise schedule was applied with a maximum times step of 100 and parameter ${\sigma_{\max }}$ set to 50. We used Adam \cite{kinga2015method} optimizer with parameters ${\beta _1} = 0.9,{\beta _2} = 0.99$. The total training steps were fixed at 700 thousand, with the initial learning rate set to ${10^{ - 4}}$ and decayed by half every 200 thousand iterations. There were 7440 data pairs in total, and the training took about 2 days. 

We compared the proposed algorithm with three different methods, including 1) U-Net-based methods: 3DU-Net-based 3D MRI motion artifact removal (3DU-Net) \cite{cciccek20163d}; 2) GAN-based methods: bootstrap subsampling and aggregation method (BSA) \cite{oh2021unpaired}; 3) Diffusion-based methods: This type of method removes motion artifacts by an iterative diffusion process and frequency domain data consistency, and we selected two typical algorithms MRI-SDE \cite{oh2023annealed} and PFAD \cite{xu2025motion}. We set the hyperparameters of the compared DL-based methods according to the original paper or official open-source codes. The training of MRI-SDE was performed following \cite{songscore}. PFAD used guided diffusion \cite{dhariwal2021diffusion} for training. 

We conducted both subjective and objective evaluations from the sagittal plane, coronal plane and axial plane. Subjective evaluation assessed the effectiveness of artifact removal and the degree of detail retention in images generated by different methods, based on visual inspection. Objective evaluation employed the commonly used peak signal-to-noise ratio (PSNR) and structural similarity index (SSIM). In addition, we adopted a new objective evaluation metric, learned perceptual image patch similarity (LPIPS) \cite{zhang2018unreasonable}, which has been proven effective for evaluating medical images \cite{kim2024systematic}.

\begin{table*}
\centering
\caption{Quantitative evaluation results of various algorithms in sagittal, axial, and coronal planes under mild motion artifacts. The best results are highlighted in bold.}
\begin{tabular}{cccccccccc} 
\hline
\multirow{2}{*}{Method} & \multicolumn{3}{c}{Coronal} & \multicolumn{3}{c}{Axial} & \multicolumn{3}{c}{Sagittal}  \\
\cline{2-10}
        &PSNR  & SSIM & LPIPS  & PSNR & SSIM & LPIPS & PSNR &SSIM  & LPIPS    \\ 
\hline
 LQ    & 32.09±1.19&0.87±0.02  &0.0435±0.0299 &32.92±0.65 &0.89±0.02 & 0.0333±0.0073             & 32.11±1.61 & 0.88±0.06 & 0.0453±0.0299                \\
 3DU-Net  &31.39±1.65  & 0.90±0.01 & 0.0788±0.0114 & 32.95±1.28 &0.94±0.01&0.0410±0.0084
     & 31.86±0.98 &  0.94±0.01 &0.0453±0.0078  \\
 BSA    &30.36±2.23  & 0.89±0.04  &  0.0463±0.0163 &30.24±1.08  & 0.87±0.01 & 0.0386±0.0063
  &30.86±0.48  & 0.89±0.01 &   0.0387±0.0063          \\
 SDE-MRI  &30.16±0.32  &0.86±0.01  & 0.0543±0.0042   &31.26±0.29  & 0.90±0.04 &0.0400±0.0066   & 30.74±0.16 & 0.92±0.01 &  0.0280±0.0032               \\
 PFAD     & 26.96±1.40 &0.75±0.02   & 0.1188±0.0143     &27.91±0.84  &0.79±0.01& 0.1064±0.0118               &28.53±1.14  & 0.83±0.06 & 0.0746±0.0552       \\
  Ours     &\textbf{34.24±1.09}  &\textbf{0.96±0.01}   &\textbf{0.0183±0.0042}                 &\textbf{35.20±1.08}  &\textbf{0.96±0.01}  &\textbf{0.0153±0.0046}               &\textbf{34.25±2.39}  &\textbf{0.96±0.00}  &\textbf{0.0181±0.0027} \\
\hline
\label{table1}
\end{tabular}
\end{table*}

\subsection{Performance Comparison on Simulated Data}
In this subsection, we evaluated the effect of motion artifact removal on mild (${M_{\min }} = 30\% ,{M_{\max }} = 45\% $) and severe (${M_{\min }} = 45\% ,{M_{\max }} = 50\% $) motion artifacts on IXI data. 

\subsubsection{Evaluation on mild motion artifacts}
Fig. \ref{fig5} shows representative results of three planes of data with mild motion artifacts recovered by different methods. We selected the yellow and green regions of interest (ROIs). Although the 3DU-Net method could effectively remove motion artifacts, the restored image exhibited significant over-smoothing issues, which may lead to the loss of some important clinical details.In contrast, the problem of detail over-smoothing in the BSA method was significantly alleviated. However, its results exhibited certain structural deformations or disappearances, as indicated by the red and yellow arrows. The method based on the diffusion model achieved favorable performance in removing motion artifacts and preserving details. The SDE-MRI method shares similar principles with the PFAD method. The two methods assume that motion artifacts occur at high frequencies. Therefore, they preserved the low-frequency k-space information of the motion-corrupted image and performed a weighted fusion of the high-frequency information with that of the motion-free image. Motion artifacts were clearly reduced, but direct weighted fusion strategy may introduce some artifacts, as indicated by the red and yellow arrows. A more serious problem was that since both models were trained on the sagittal plane, there were inter-slice discontinuities in the axial and coronal planes. These discontinuities may hinder clinical diagnosis. In comparison, our method was able to better remove motion artifacts and was the most structurally similar to the ground truth. Notably, since we combined information from different planes, we achieved superior results in resolving inter-slice discontinuities.


Table \ref {table1} presents the quantitative results of all methods. Our method significantly outperforms the others. All 2D-based comparison algorithms were trained on the sagittal plane, resulting in significantly inferior performance in other planes. In contrast, our algorithm achieves favorable performance across all planes.

\subsubsection{Evaluation on severe motion artifacts}
Fig. \ref{fig6} shows representative results of three plans of data with severe motion artifacts recovered by different methods.
Due to severe motion artifacts, brain tissue structures captured in MRI images were no longer useful for clinical diagnosis. The 3DU-Net method was nearly ineffective in the presence of severe motion artifacts. The BSA method achieved superior artifact suppression compared to 3DU-Net, but exhibited significant structural inaccuracies. The SDE-MRI method was ineffective in removing severe motion artifacts. We speculate that this is because the image generated by directly weighting and fusing motion-corrupted and motion-free images in the Fourier domain is significantly affected by motion artifacts, resulting in substantial motion artifacts in the generated image. The PFAD method, on the other hand, performed fusion not only in the Fourier domain but also in the image domain, which may account for its superior artifact removal. However, this method also introduced significant structural distortions while removing artifacts. Overall, these comparison methods struggled to balance artifact removal and structure preservation in the presence of severe motion artifacts. In contrast, our method enables the diffusion model to learn the degradation relationship between motion-corrupted and motion-free images, enabling both artifact removal and structure preservation.

Table \ref{table2} presents the quantitative results of all methods. Our method also significantly outperformed the others. Especially for the LPIPS metric, our method outperforms the comparison algorithm by almost an order of magnitude.

\begin{table*}
\centering
\caption{Quantitative evaluation results of various algorithms in sagittal, axial, and coronal planes under severe motion artifacts. The best results are highlighted in bold.}
\begin{tabular}{cccccccccc} 
\hline
\multirow{2}{*}{Method} & \multicolumn{3}{c}{Coronal} & \multicolumn{3}{c}{Axial} & \multicolumn{3}{c}{Sagittal}  \\
\cline{2-10}
        &PSNR  & SSIM & LPIPS  & PSNR & SSIM & LPIPS & PSNR &SSIM  & LPIPS    \\ 
\hline
 LQ    & 21.35±0.94 &0.61±0.04  &0.2162±0.0289 &22.38±0.71 &0.71±0.03 & 0.1352±0.0211   & 21.57±2.23 & 0.54±0.04 & 0.2460±0.0219                \\
 3DU-Net  &21.38±0.61  & 0.59±0.02 & 0.2149±0.0184 & 22.53±1.03 &0.64±0.02&0.1611±0.0176 & 21.84±2.72 & 0.57±0.01 &0.1788±0.0421  \\
 BSA    &24.87±0.90  & 0.74±0.04  &  0.1213±0.0253 &25.94±0.78  & 0.82±0.01 &0.0755±0.0105
  &25.22±3.32  &0.72±0.07 &   0.1311±0.0182          \\
 SDE-MRI  &22.01±0.71 &0.64±0.02  & 0.1992±0.0210   &23.01±0.47 &0.72±0.03 &0.1355±0.0174   &21.97±2.35& 0.56±0.06 &  0.2105±0.0189               \\
 PFAD     & 21.88±0.85 &0.59±0.04  & 0.2121±0.0321    &22.91±0.85  &0.68±0.01& 0.1476±0.0146    &22.56±2.04 & 0.53±0.04 & 0.1938±0.0325      \\
  Ours     &\textbf{28.74±1.05}  &\textbf{0.86±0.05}   &\textbf{0.0549±0.0102}                 &\textbf{30.39±1.56}  &\textbf{0.90±0.04}  &\textbf{0.0433±0.0111}               &\textbf{29.36±3.85}  &\textbf{0.85±0.06}  &\textbf{0.0691±0.0223} \\
\hline
\label{table2}
\end{tabular}
\end{table*}

\subsection{Performance Comparison on Actual Data}
\begin{figure*}
\centerline{\includegraphics[width=\textwidth]{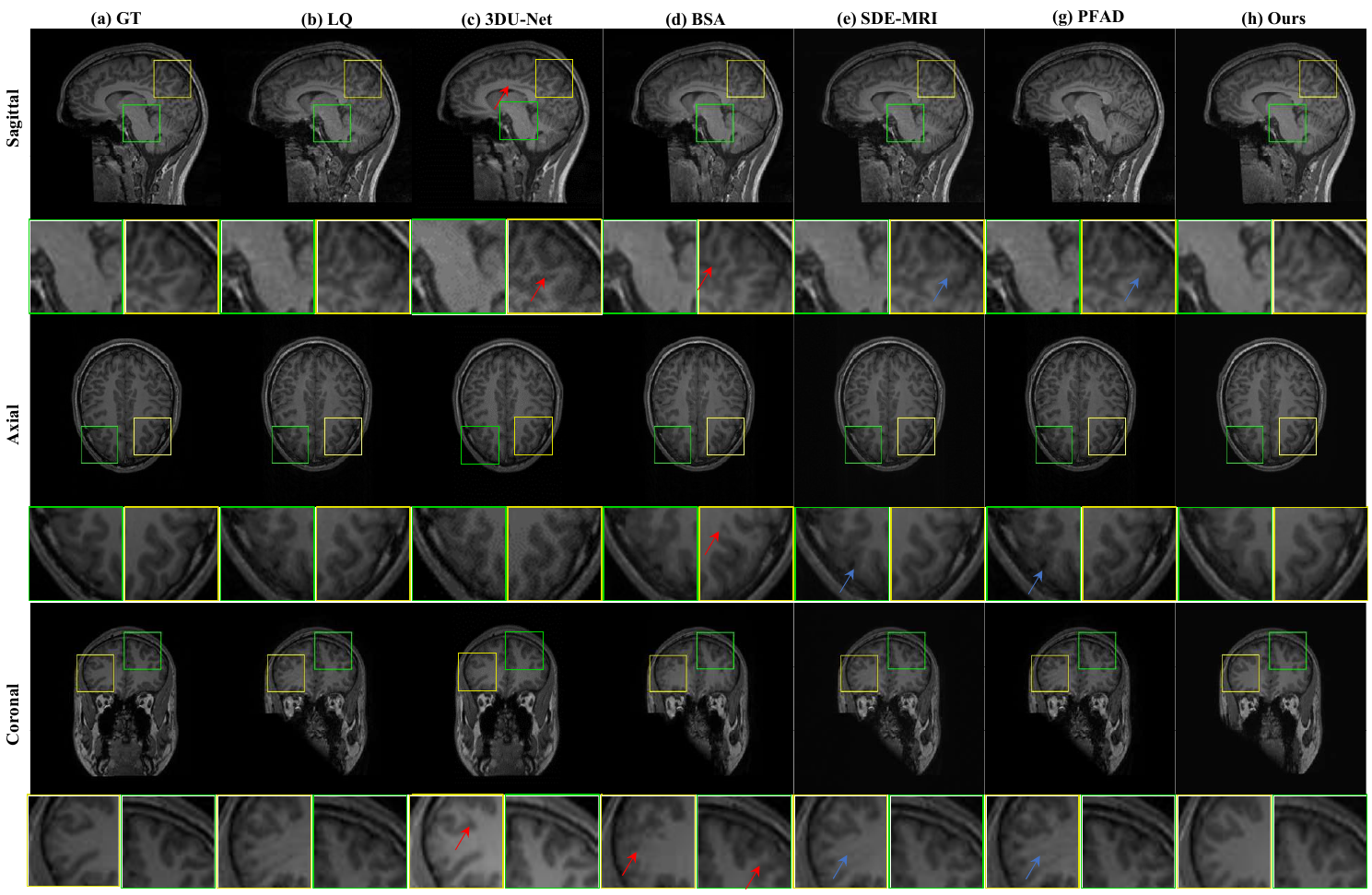}}
\caption{Qualitative results of real world motion artifact correction for brain data in sagittal, coronal, and horizontal plane views using various methods. Yellow and green ROIs are presented as local magnifications. The first column denotes the ground truth (GT), the second column denotes the low-quality image (LQ) corrupted by motion artifacts, the third to sixth columns denote results of comparison algorithms, and the last column denotes our proposed method.}
\label{fig7}
\end{figure*}
To verify the effectiveness of the algorithm in the real world, we deployed our algorithm and the comparison algorithms on real world data. The results are shown in Fig. \ref{fig7}. The 3DU-Net method benefits from paired learning and is effective at removing artifacts. However, current real-world data is not fully paired, causing the method to recover spurious structures that do not belong to the LQ images, which are most noticeable in the coronal plane. Furthermore, it exhibits significant over-smoothing compared to GT images. The BSA method is significantly more faithful to the LQ structure than 3DU-Net, but some errors in the results are also present, as shown by the red arrows. The SDE-MRI method can basically preserve the structure of the LQ images, but the regions indicated by pink arrows show insufficient artifact removal. This may be due to the inherent flaws of the method's direct employment of weighted fusion. The PFAD method is better at removing artifacts than SDE-MRI, but significant structural errors and inter-slice discontinuities are observed, as indicated by the red arrows. This also reflects the imperfections of the two diffusion model-based methods in terms of structure preservation and artifact removal. In contrast, our method exhibits the capability to remove artifacts while preserving details in the LQ images even with unpaired data. This is due to the fact that our method does not learn the correspondence between the LQ images and the ground-truth images, but rather learns the degradation process from an image with motion artifacts to one without such artifacts. Furthermore, our pseudo-3D learning strategy avoids the inter-layer discontinuities that arise in 2D learning. 

We randomly selected ROIs and evaluated the PSNR and SSIM metrics of several algorithms, as shown in Fig. \ref{fig7}. Because the 3DU-Net method completely learns the characteristics of the ground truth (GT) and fails to retain the key features of the LQ images, we believe that comparing its objective evaluation metrics against the GT is meaningless. Finally, our algorithm also exhibits significant advantages in terms of objective evaluation metrics.

\subsection{Temporal evaluation of 3D motion artifact removal}
\begin{table}
\centering
\caption{The average time required for different diffusion-based methods to remove motion artifacts of a $240 \times240 \times240 $ 3D volume was calculated.}
\begin{tabular}{ccccc} 
\hline
 &SDE-MRI  &PFAD  &w/o Wav  &w Wav(Ours)   \\ 
\hline
time(min) &1184  &80  &35  &\textbf{14}   \\
\hline
\label{table3}
\end{tabular}
\end{table}
Diffusion-based methods are often challenging to deploy due to efficiency issues. Fortunately, our method employs a mean-reverting strategy to significantly reduce the number of sampling steps required. Furthermore, we utilize a wavelet transform to perform dimensionality reduction on the input data, further accelerating the sampling process. As shown in Table \ref{table3}, the mean-reverting strategy reduces the sampling time to 35 minutes, while the wavelet transform reduces the dimensionality to 40\% of the original.
\subsection{Ablation studies}
\subsubsection{Effectiveness of Wavelet Residual Block}

\begin{figure}[!t]
\centerline{\includegraphics[width=\linewidth]{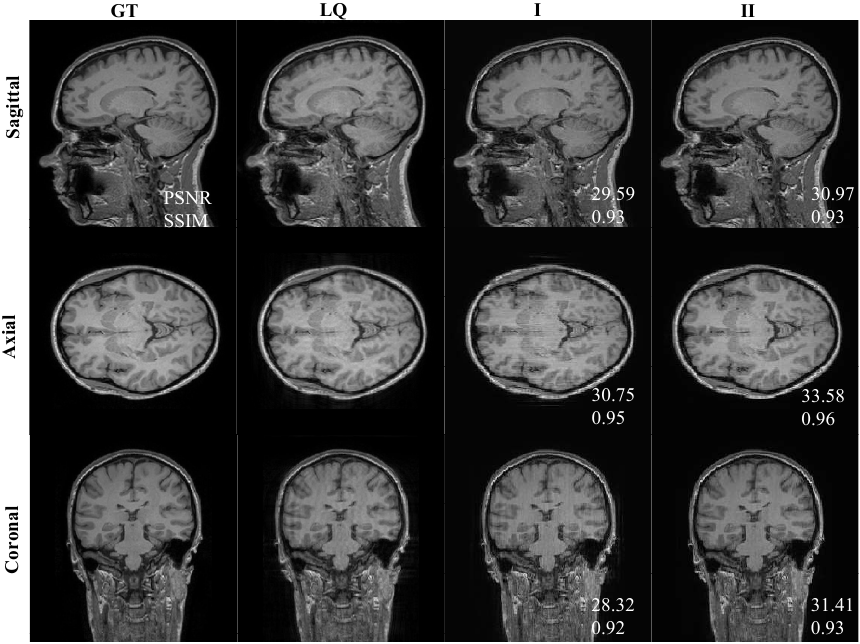}}
\caption{Ablation study of wavelet residual blocks. The first row shows the coronal plane, the second row shows the horizontal plane, and the third row shows the sagittal plane. \Rmnum{1} shows the 2D motion artifact removal result using the original residual block model, and \Rmnum{2} shows the 2D motion artifact removal result using the wavelet residual block model.}
\label{fig8}
\end{figure}

To validate the effectiveness of the wavelet residual block, we conducted the following ablation experiment. Specifically, we employed the original IR-SDE model as a baseline. We subsequently replaced the residual block in the original IR-SDE model with our proposed wavelet residual block. As shown in Fig. \ref{fig8}, the model using the wavelet residual block exhibited significant improvements in visual evaluation and objective metrics. This is because the convolution layer in our wavelet residual block uses a wavelet convolution module. This convolutional module achieves a larger receptive field and more effectively captures low-frequency information within MRI images.

\subsubsection{Effectiveness of Pseudo-3D Strategies}

\begin{figure}[!t]
\centerline{\includegraphics[width=\linewidth]{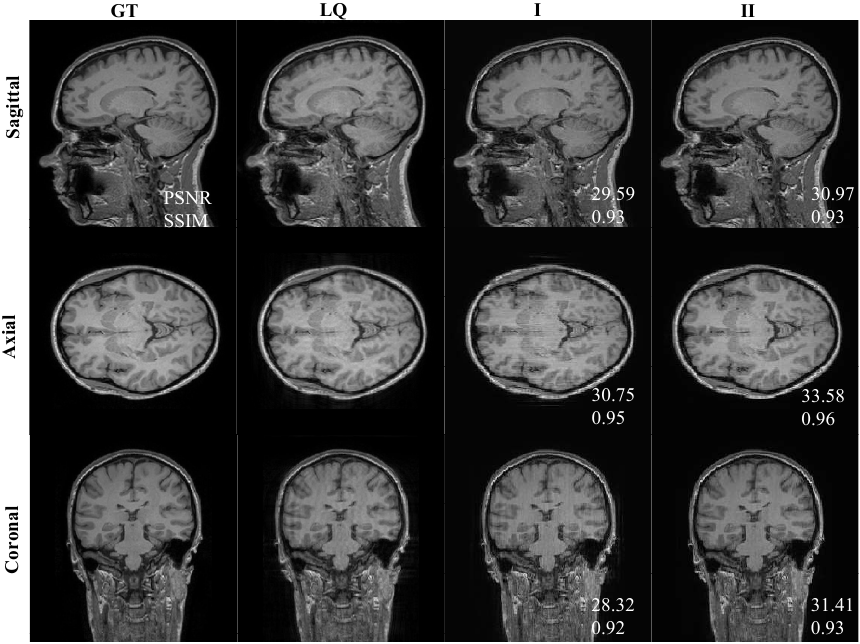}}
\caption{Ablation study of pseudo-3D strategies. The first row shows the coronal plane, the second row shows the horizontal plane, and the third row shows the sagittal plane. \Rmnum{3} means using 2D strategy, and \Rmnum{4} means using pseudo 3D strategy.}
\label{fig9}
\end{figure}

We call this approach of approximating a 3D distribution using two orthogonal 2D distributions as a pseudo-3D strategy. To more effectively validate the effectiveness of the pseudo-3D strategy, we conducted the following ablation studies on severe motion-corrupted MRI images. Our baseline consists of the original IR-SDE and the wavelet residual block. Based on this, we trained and tested the model using the proposed pseudo-3D strategy. The results demonstrate that the proposed method can significantly eliminate artifacts caused by 3D discontinuities, as shown in Fig. \ref{fig9}. Notably, adopting this strategy not only enhanced performance in the baseline's non-training planes (axial and coronal planes), but also improved the effects in the baseline's training plane (sagittal plane). This demonstrates that although our method is still trained in 2D, the proposed pseudo-3D strategy can effectively capture the prior distribution of 3D data. 
\section{Discussion}
\begin{figure}[!t]
\centerline{\includegraphics[width=\linewidth]{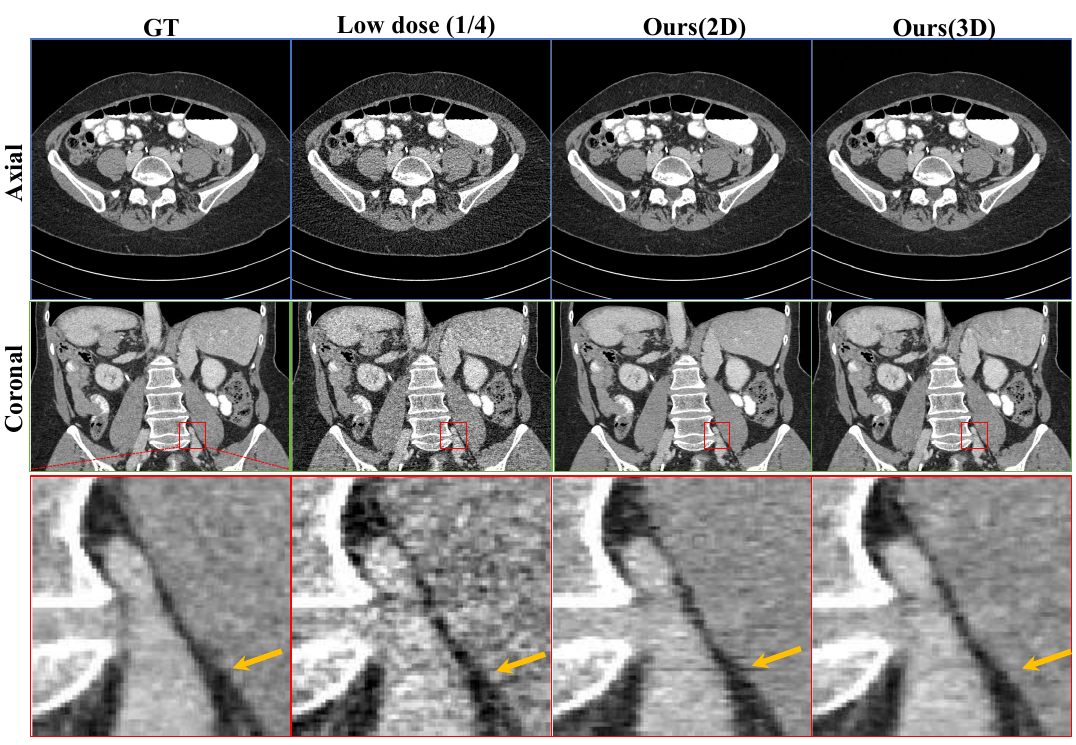}}
\caption{The performance of our algorithm in denoising CT images. The red area is the ROI region on the coronal plane.}
\label{fig10}
\end{figure}

\subsection{Potential for application to various tasks}
The algorithm proposed in this study demonstrated superior performance in the 3D MRI motion artifact removal task, effectively enhancing the quality of artifact-affected images and demonstrating its practicality in this specific scenario. Notably, the algorithm's core design principle utilizes two pre-trained 2D data distributions to approximate a 3D data distribution and employs a mean-reverting SDE to model the degradation process from high-quality 3D volumes to low-quality 3D volumes. Crucially, our approach does not require in-depth modeling of the degradation process. Taking all the above features into consideration, this approach is not limited to motion artifact degradation alone. Its deep exploration of the inherent spatial continuity and structural correlations of 3D medical images, as well as its ability to discern and repair degraded information, offer potential for extension to other 3D medical image restoration tasks. For example, in the 3D image denoising task, where noise blurs details or distorts signals, the algorithm employs a similar 3D feature extraction and restoration framework to effectively suppress noise while preserving key anatomical information. As shown in Fig. \ref{fig10}, our algorithm was deployed on data provided by the AAPM Low Dose Challenge \cite{moen2021low}, aiming to remove noise from 1/4 dose CT images. The proposed algorithm (under 2D conditions) shows good denoising performance, but inevitably suffers from inter-slice discontinuity issues, as shown by the yellow arrows. The discontinuity problem is eliminated after integrating our pseudo-3D strategy. This potential not only demonstrates the versatility of algorithm design, but also provides new insights for solving a wider range of 3D medical image quality problems.

\subsection{Limitations of the proposed model}
We acknowledge certain limitations of this work. First, the proposed method does not require explicit modeling of the degradation process, but still relies on paired data for training. Although our method performs well on weakly paired data, its performance degrades compared to when using fully paired data. In the future, we plan to aim to reduce the reliance on paired data by introducing cycle consistency losses \cite{zhu2017unpaired} or employing ideas such as learning from further corrupted images \cite{daras2023ambient}. 
Secondly, although the proposed model is much faster than other diffusion-based motion artifact removal methods, it is remains slower than GAN-based and U-Net-based methods. In the future, we plan to incorporate methods such as latent diffusion techniques \cite{rombach2022high}, consistency models \cite{song2023consistency}, or flow matching \cite{lipman2023flow} to accelerate our method.

\section{Conclusion}
In this work, we propose an end-to-end wavelet-optimized framework (3D-WMoCo) for 3D MRI motion correction using pre-trained 2D score priors. This approach (i) leverages two orthogonal pre-trained 2D scores to drive the 3D distribution prior and learns the degradation process of 3D data via a mean-reverting stochastic differential equation; (ii) incorporates wavelet diffusion to further reduce inference time; and (iii) designs a wavelet residual block to expand the receptive field. Experimental validation results demonstrate that the proposed 3D-WMoCo method exhibits strong effectiveness in both simulated and real-world motion artifact tests. Notably, our method is a general approach well-adapted for scenarios where the degradation process is difficult to model explicitly.






\bibliographystyle{IEEEtran}
\bibliography{ref}

\vfill

\end{document}